\documentstyle[multicol,aps,psfig]{revtex}
\renewcommand{\narrowtext}{\begin{multicols}{2}
\global\columnwidth20.5pc\noindent}
\renewcommand{\widetext}{\end{multicols}
\global\columnwidth42.5pc}
\begin{document}
\draft
\preprint{31 December 2001}
\title{Re-Entrant Quantum Phase Transitions in Antiferromagnetic
       Spin-1 Ladders}
\author{Shoji Yamamoto}
\address
{Department of Physics, Okayama University,
 Tsushima, Okayama 700-8530, Japan}
\author{T$\hat{\mbox o}$ru Sakai}
\address
{Department of Electronics, Tokyo Metropolitan Institute of
 Technology, Hino, Tokyo 191-0065, Japan}
\author{Akihisa Koga}
\address
{Department of Applied Physics, Osaka University,
 Suita, Osaka 565-0871, Japan}

%\date{Received \hspace{6cm}}
\date{Received 31 December 2001}
\maketitle
\begin{abstract}
In response to recent chemical attempts to construct higher-spin
ladder materials from organic polyradicals, we study the ground-state
properties of a wide class of antiferromagnetic spin-$1$ ladders.
Employing various numerical tools, we reveal the rich phase diagram
and {\it correct a preceding nonlinear-sigma-model prediction}.
A variational analysis well interprets the phase competition with
particular emphasis on the {\it re-entrant phase boundary} as a
function of the rung interaction.
\end{abstract}
\narrowtext

   In 1983 Haldane \cite{H464} awoke renewed interest in quantum spin
chains predicting a striking contrast between integer- and
half-odd-integer-spin Heisenberg antiferromagnets.
His argument was indeed verified in a spin-$1$ material
Ni(C$_2$H$_8$N$_2$)$_2$NO$_2$(ClO$_4$) \cite{R945}
and was given an analytic support \cite{A799} as well.
Since then the energy gaps in magnetic excitation spectra, that is,
spin gaps, have been a central issue in materials science.
In the last decade more and more researchers made a wide variety of
explorations into the spin-gap problem, such as the spin-Peierls
transition in inorganic compounds \cite{H3651}, quantized plateaux in
magnetization curves \cite{O1984}, and antiferromagnetic gaps in the
ferromagnetic background \cite{Y14008}.
Among others Dagotto {\it et al.} \cite{D5744} pointed out that
another mechanism of the gap formation should lie in a ladder$-$two
coupled chains.
A spin gap was indeed observed in a typical two-leg ladder material
SrCu$_2$O$_3$ \cite{A3463}.
Moreover, superconductivity was brought about in its hole-doped
version (SrCa)$_{14}$Cu$_{24}$O$_{41}$ \cite{U2764}.
Ladder systems caused us further surprise exhibiting excitation
spectra varying with the number of their legs \cite{D618}.

   So far metal oxides have been representative of ladder materials.
Though molecule-based ones \cite{H12649,L100402} have been
synthesized in an attempt to reduce the spin gaps and obtain
experimental access to them, the situation of copper ions supplying
the relevant spins remains unchanged.
Therefore they are all spin-$\frac{1}{2}$ antiferromagnets.
In such circumstances there has occurred a brand-new idea of
constructing purely organic ladder systems.
Katoh {\it et al.} \cite{K1008} synthesized novel organic biradicals
and tetraradicals which crystallize to form an antiferromagnetic
ladder of spin-$\frac{1}{2}$ and that of effectively spin-$1$,
respectively.
Their polyradical strategy has yielded further harvest such as an
effective spin-$1$ antiferromagnet on a honeycomb lattice
\cite{H12924} and a ladder ferrimagnet of mixed spins $1$ and
$\frac{1}{2}$ \cite{H}, displaying the wide tunability of the
crystalline structures in higher-spin systems as well.

   Distinct spin-gap mechanisms may lie in higher-spin ladders and
quantum competition between them must lead us to further enthusiasm
for ladder systems.
There exist pioneering works in this context.
Sierra \cite{S3299} generalized the well-known nonlinear-sigma-model
analysis \cite{H464,A397} on quantum spin chains to multi-leg ladder
systems.
His findings suggest that only an odd number of half-odd-integer-spin
chains are massless, supporting experimental observations on a series
of spin-$\frac{1}{2}$ ladder antiferromagnets \cite{A3463,E2626}.
The technique was further developed for spatially inhomogeneous
ladders \cite{M3443}.
Mixed-spin ladders \cite{F398,L343} were also investigated with
particular emphasis on the competition between massive and massless
phases.

   In comparison with extensive calculations
\cite{D5744,P4393,W886,F3714} on spin-$\frac{1}{2}$ ladders,
there exist few quantitative investigations into higher-spin ladders
\cite{T} and most of the above-mentioned predictive theories remain
to be verified.
Hence, in this article, we solve the ground-state properties of
antiferromagnetic spin-$1$ ladders with two legs.
Employing various numerical tools and complementing them by a
variational argument, we elucidate the {\it valence-bond-solid-like
nature} of their ground states, which is in contrast with the
spin-liquid or resonating-valence-bond ground states
\cite{P4393,W886,A1196,R445} of spin-$\frac{1}{2}$ two-leg ladders.
The obtained phase diagram is reminiscent of the preceding
sigma-model prediction \cite{M3443} but contains a {\it re-entrant
phase boundary}, which can never be extracted from any
field-theoretical argument.

   Considering that an advantage of assembling organic open-shell
molecules into a magnetic material is the isotropic intermolecular
exchange couplings, while the polyradical strategy is accompanied by
spatial variations in magnetic interaction \cite{H12924}, we treat
a wide class of spin-$1$ antiferromagnetic ladders
\begin{equation}
   {\cal H}
   =\sum_{j=1}^L
    \left(
     \sum_{i=1}^2
      J_{\parallel}\gamma_{i,j}
      \mbox{\boldmath$S$}_{i,j}\cdot\mbox{\boldmath$S$}_{i,j+1}
    + J_{\perp}
      \mbox{\boldmath$S$}_{1,j}\cdot\mbox{\boldmath$S$}_{2,j}
    \right)\,,
   \label{E:H}
\end{equation}
\begin{figure}
\centerline
{\qquad\qquad\ \mbox{\psfig{figure=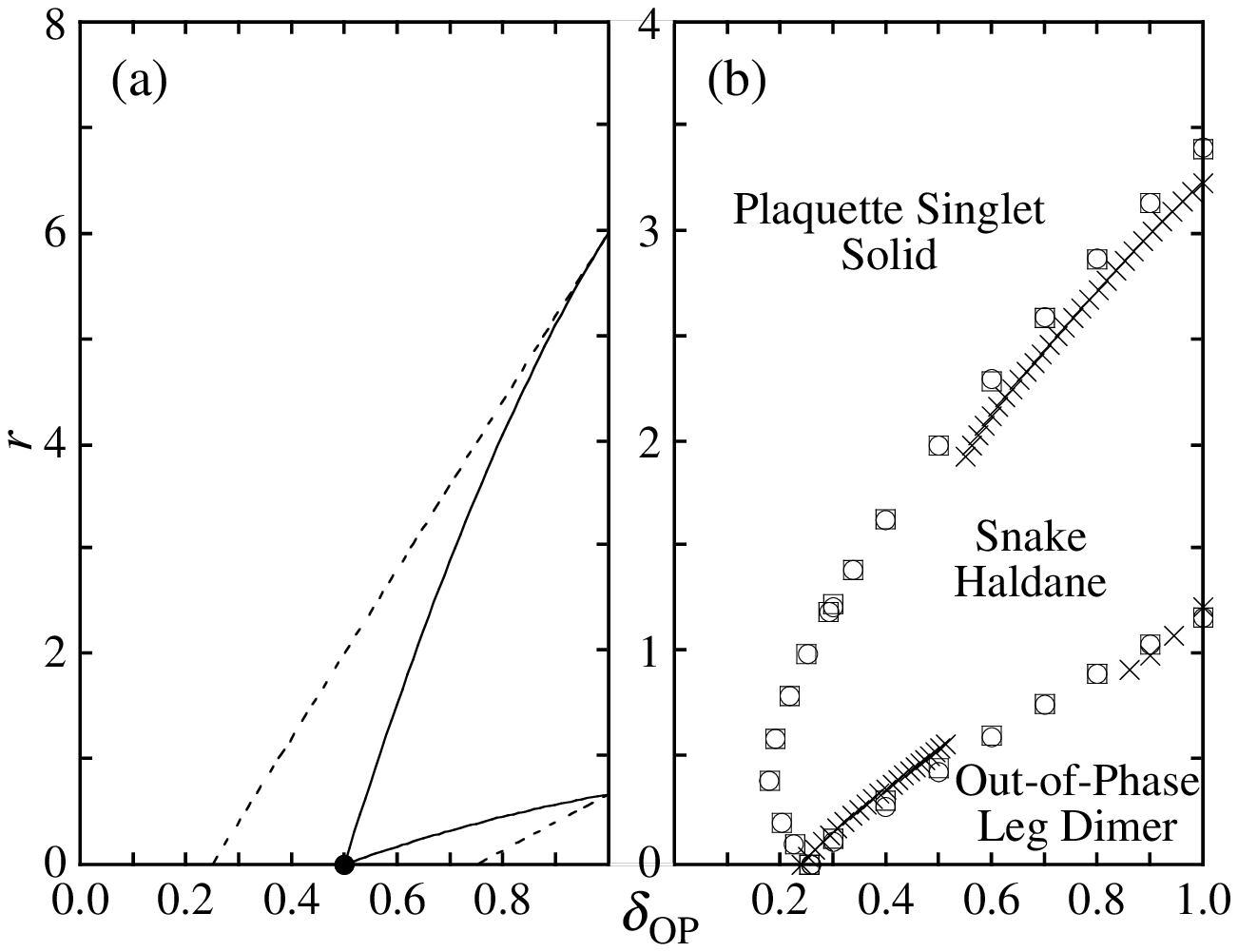,width=90mm,angle=0}}}
\vskip 3mm
\caption{Phase diagrams for the antiferromagnetic spin-$1$ ladder
         with two out-of-phase legs.
         (a) A field-theoretical prediction [19].
         The two critical lines (dashed lines) derived from the
         effective sigma model for ladders are inconsistent with the
         sigma-model analysis on isolated chains ($\bullet$).
         They remain far apart from each other even in the
         decoupled-chain limit $r=0$.
         Therefore qualitatively patched-up phase boundaries (solid
         lines) were predicted.
         (b) Our numerical findings.
         The series-expansion estimates are shown by $\times$, while
         the level-spectroscopy analyses by $\Box$ ($L=6$) and
         $\circ$ ($L=8$).}
\label{F:PhD}
\vspace*{-4mm}
\centerline
{\qquad\qquad\mbox{\psfig{figure=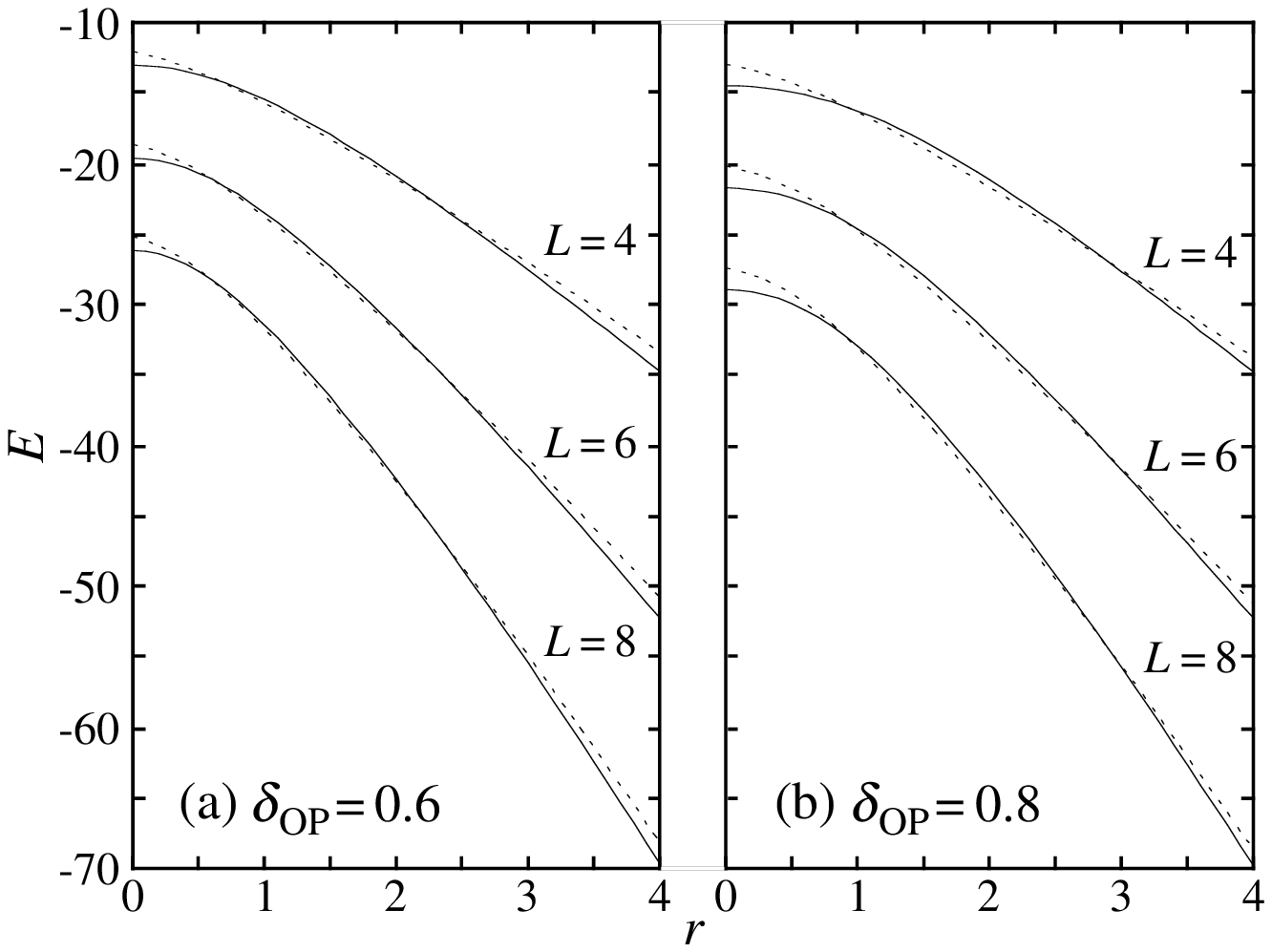,width=94mm,angle=0}}}
\vskip 2mm
\caption{Demonstration of the level spectroscopy.
         The lowest-lying two eigenvalues in the subspace of zero
         magnetization as functions of $r$ cross at transition points
         provided the twisted boundary condition is imposed on the
         Hamiltonian.}
\label{F:level}
\end{figure}
\vskip 1mm
\noindent
where the bond-alternation parameter $\gamma_{i,j}$ is defined in two
ways as
\begin{equation}
   \gamma_{i,j}=
   \left\{
    \begin{array}{ll}
     1+(-1)^{i+j}\delta_{\rm OP}\ \ &\mbox{(out-of-phase legs)}\,,\\
     1+(-1)^{  j}\delta_{\rm IP}\ \ &\mbox{(in-phase legs)}\,.
    \end{array}
   \right.
\end{equation}
We calculate the region of
$0\leq\delta_{\rm OP}(\delta_{\rm IP})\leq 1$ and hereafter set
$J_{\perp}/J_{\parallel}$ to $r(\geq 0)$.
Martin-Delgado, Shankar, and Sierra \cite{M3443} studied the cases of
out-of-phase legs deriving a low-energy-relevant sigma model.
For the spin-$S$ ladders with two out-of-phase legs, the topological
angle in the effective sigma model turns out
$8\pi S\delta_{\rm OP}/(r+2)$ and reads as the critical lines
$8S\delta_{\rm OP}=(2n+1)(r+2)$
$(n=0,\pm 1,\cdots)$.
However, these findings do not smoothly merge with the
well-established critical behavior in one dimension,
$2S(1-\delta)=2n+1$ \cite{A397}, as is shown in Fig. \ref{F:PhD}(a).
Thus, it is necessary to verify the true scenario all the more in
higher dimensions.

   One of the most reliable solution may be a numerical analysis
\cite{S4053} on the phenomenological renormalization-group equation
\cite{N561}.
However, the scaled gaps are ill-natured due to the close critical
points, so as to make the fixed points hard to extract from
available numerical data.
Then we switch our strategy to the level spectroscopy \cite{N5773},
the core idea of which is summarized as detecting transition points
by crossing of two relevant energy levels.
Although the method is generically applicable to the Gaussian
critical points \cite{Y16128}, no explicit change of symmetry
accompanies the present phase transitions and therefore any levels
do not cross naively.
In order to overcome the difficulty of this kind, Kitazawa
\cite{K285} proposed the idea of applying the twisted boundary
condition, that is to say in the present case, setting the boundary
exchange couplings equal to
$-\sum_{i=1}^2 J_\parallel \gamma_{i,L}
(S_{i,L}^xS_{i,1}^x+S_{i,L}^yS_{i,1}^y-S_{i,L}^zS_{i,1}^z)$.
Then the energy structure of the Hamiltonian is changed and the
lowest two levels are led to cross at transition points, which is
demonstrated in Fig. \ref{F:level}.
Due to the limit of time and memory well spent, we have restricted
our calculations up to $L=8$.
We plot in Fig. \ref{F:PhD}(b) bare findings for the crossing points
at $L=6$ and $L=8$ rather than extrapolate them trickily.
We are sure that the data uncertainty still left is within the symbol
size.
A series-expansion technique \cite{S2484,K6133} guarantees the level
spectroscopy to work well.
Starting with decoupled singlet dimers on legs or rungs and
expanding the energy gap as a power series in a relevant perturbation
parameter, we can obtain a partial knowledge of phase transitions.
Here we have calculated the gap up to the ninth order and further
applied the Dlog Pad\'e approximants \cite{G13} to them.
The thus-obtained phase boundaries, which are also shown in Fig.
\ref{F:PhD}(b), elucidate the nature of the phase competition, that
is, the Affleck-Kennedy-Lieb-Tasaki (AKLT) valence-bond-solid (VBS)
\cite{A799} on a snakelike path versus decoupled dimers.

\begin{figure}
\centerline
{\qquad\qquad\qquad\ 
 \mbox{\psfig{figure=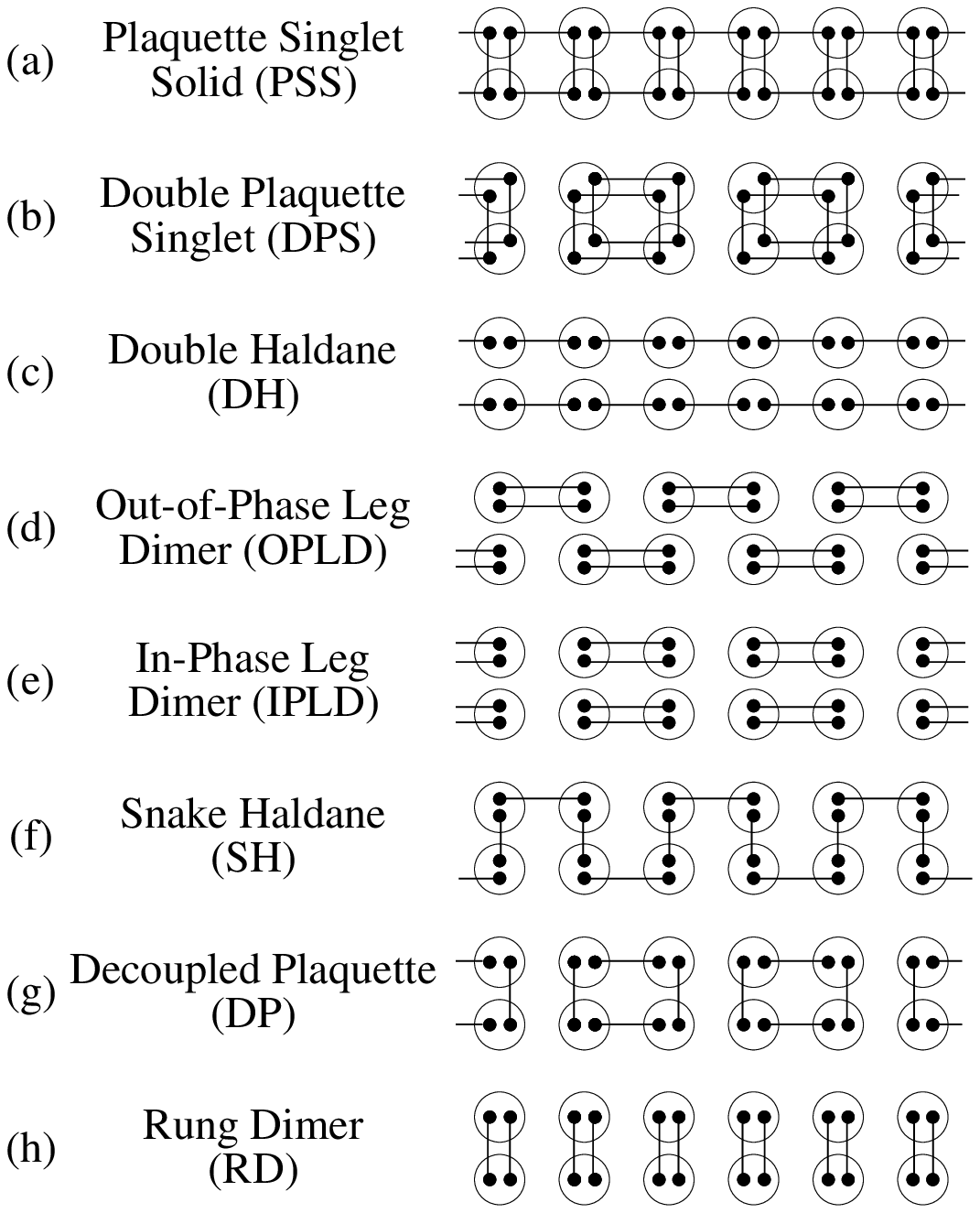,width=78mm,angle=0}}}
\vspace*{4mm}
\caption{Plaquette-singlet-solid and valence-bond-solid states
         relevant to the two-leg antiferromagnetic spin-$1$ ladders.
         $\bullet$ denotes a spin $\frac{1}{2}$ and their
         segment linkage means a singlet formation.
         $\bigcirc$ represents an operation of constructing a spin
         $1$ by symmetrizing the two spin $\frac{1}{2}$'s inside.}
\label{F:VBS}
\vspace*{6mm}
\centerline
{\mbox{\psfig{figure=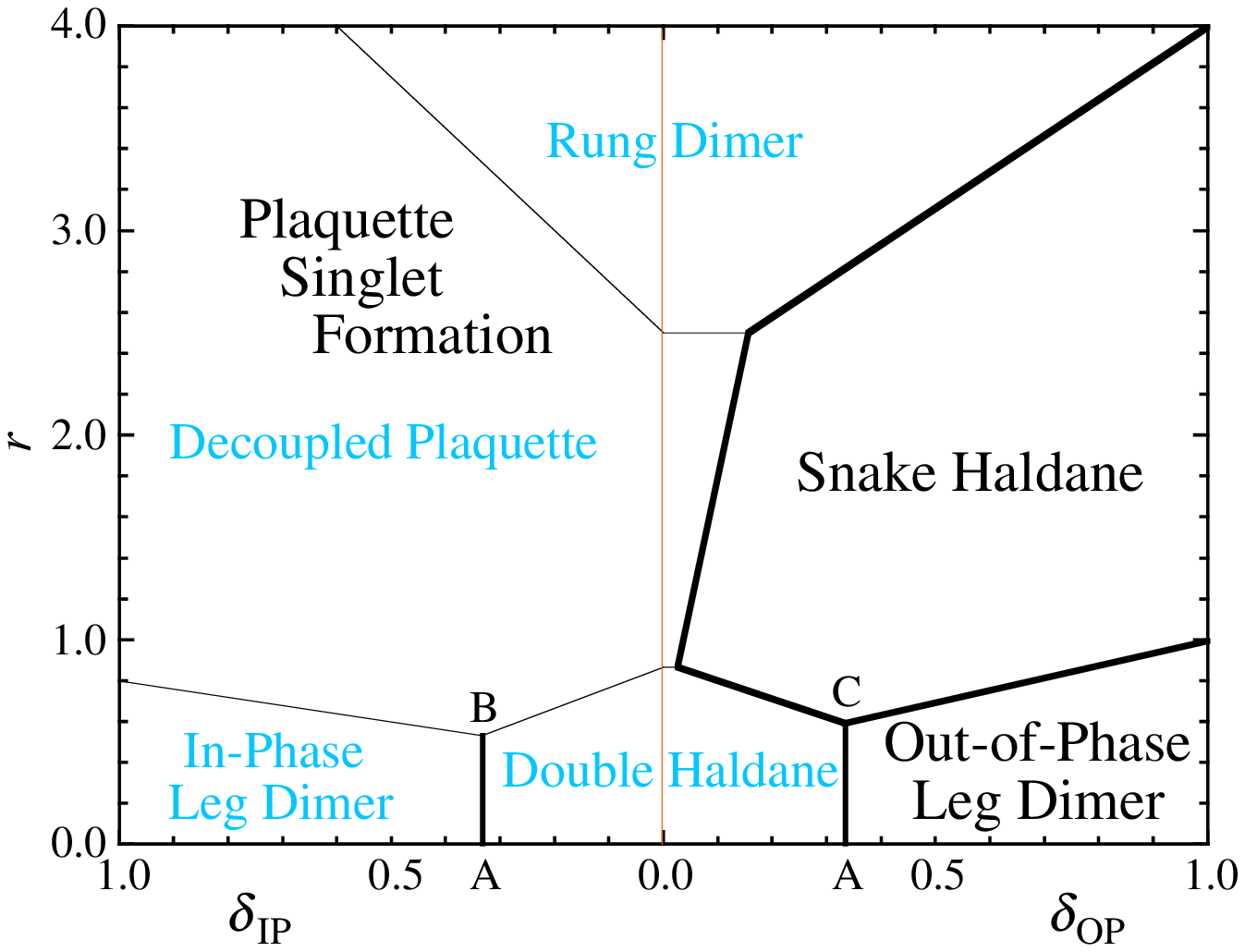,width=100mm,angle=0}}\quad}
\vspace*{-57mm}
\caption{Variational phase diagrams for the two-leg antiferromagnetic
         spin-$1$ ladders.
         The thick solid lines describe phase transitions, whereas
         the thin ones represent the crossover of the ground-state
         nature within the present variational scheme.
         The dotted line is only a guide for eyes.}
\label{F:PhDval}
\end{figure}

   The most impressive findings are {\it re-entrant quantum phase
transitions} with increasing $r$.
The preceding sigma-model analysis \cite{M3443} is indeed
enlightening but never able to reveal this novel quantum behavior.
In order to characterize each phase, let us consider a variational
approach.
We know that singlet dimers on rungs [Fig. \ref{F:VBS}(h)] are
stabilized for $r\rightarrow\infty$, whereas either dimers on legs
[Figs. \ref{F:VBS}(d) and \ref{F:VBS}(e)] or the double AKLT VBS
[Fig. \ref{F:VBS}(c)] for $r\rightarrow 0$.
Two more interchain VBS states [Figs. \ref{F:VBS}(f) and
\ref{F:VBS}(g)] may be adopted as variational components for the
intermediate-$r$ region.
Thus the linear combination of Figs. \ref{F:VBS}(c) to \ref{F:VBS}(h)
can be an approximate ground-state wave function for spin-$1$
ladders.
Since the present variational components are all asymptotically
orthogonal to each other, the variational ground state turns out
any of them itself \cite{Y3603}.
The thus-obtained phase diagram is presented in Fig. \ref{F:PhDval}.
The significant stabilization of the intermediate phase, which is now
characterized as SH, and the resultant re-entrant phase boundary are
successfully reproduced.
Considering that a couple of critical chains immediately turn massive
with their rung interaction switched on \cite{D5744}, the point C
should coincide with the point A under more refined (and thus
inevitably numerical) variational investigation.

   The present variational calculation implies possible phase
transitions for in-phase-leg ladders as well, but this is totally due
to the naive wave function.
Numerical observation of the energy structure ends up with no gapless
\quad
\vspace*{1mm}
\noindent
point in this region.
The sigma-model approach also concludes no critical point, giving
the topological angle $4\pi S$ independent of both $r$ and
$\delta_{\rm P}$.
The key to the ground-state nature of in-phase-leg ladders is the
four-spin correlation \cite{T}.
Let us consider interacting four spins of $S=\frac{1}{2}$ which are
described by the Hamiltonian
$
   {\cal H}
   = \sum_{i=1}^2
     \mbox{\boldmath$S$}_{i,l}\cdot\mbox{\boldmath$S$}_{i,l+1}
   +r\sum_{j=l}^{l+1}
     \mbox{\boldmath$S$}_{1,j}\cdot\mbox{\boldmath$S$}_{2,j}\,.
$
In terms of the Schwinger boson representation:
$S^+=a^\dagger b$;
$S^-=a b^\dagger$;
$S^z=\frac{1}{2}(a^\dagger a-b^\dagger b)$;
$\hat{S}=\frac{1}{2}(a^\dagger a+b^\dagger b)$,
their ground state is explicitly given as
\begin{eqnarray}
   &&
   |\psi(r)\rangle_l
   =\biggl[
     \cos\theta(r)
     \prod_{i=1}^2
     \left(
      a_{i,l}^\dagger b_{i,l+1}^\dagger
     -b_{i,l}^\dagger a_{i,l+1}^\dagger
     \right)
   \nonumber \\
   &&\qquad
    +\sin\theta(r)
     \prod_{j=l}^{l+1}
     \left(
      a_{1,j}^\dagger b_{2,j}^\dagger
     -b_{1,j}^\dagger a_{2,j}^\dagger
     \right)
    \biggr]
    |0\rangle_l\,,
   \label{E:PS}
\end{eqnarray}
where $|0\rangle_l$ is the Bose vacuum and $\theta(r)$ is given by
$\tan\theta(r)=r-1+\sqrt{1-r+r^2}$.
As $r$ varies from $0$ to $\infty$, $\theta$ moves from $0$ to
$\frac{\pi}{2}$, that is, {\it the leg dimers continuously turn into
the rung dimers.}
Using the {\it plaquette singlet} state (\ref{E:PS}), we can in
principle construct much better variational wave functions for
spin-$1$ ladders particularly in the in-phase-leg region as
\begin{eqnarray}
   &&
    |\Psi(r)\rangle=
    \prod_{i,j}{\cal P}_{i,j}
    \prod_l |\psi(r)\rangle_{2l-1}
   \nonumber \\
   &&\quad\times
    \Bigl[
     \cos\phi(\delta_{\rm IP})|\psi(r)\rangle_{2l-1}
    +\sin\phi(\delta_{\rm IP})|\psi(r)\rangle_{2l  }
    \Bigr]\,,
   \label{E:PSS}
\end{eqnarray}
where ${\cal P}_{i,j}$ represents an operation of symmetrizing the
two spin $\frac{1}{2}$'s at site $(i,j)$ into a spin $1$.
Now that $\theta(r)$ is a continuous function of $r$ and may here
deviate from that in Eq. (\ref{E:PS}), a naive optimization
\cite{Y3603} of Eq. (\ref{E:PSS}) is no more feasible.
However, the refined variational scheme shows us more.
For better understanding of the wave function (\ref{E:PSS}), we
visualize in Fig. \ref{F:VBS} its special forms for
$\phi=\frac{\pi}{2}$ (a) and $\phi=0$ (b), which are most stabilized
at $\delta_{\rm IP}=0$ and $\delta_{\rm IP}=1$, respectively.
The {\it snapshots} of PSS (DSP) at $\theta=0$,
$\theta=\frac{\pi}{4}$, and $\theta=\frac{\pi}{2}$ are nothing but
the variational components DH (IPLD), DP, and RD, respectively.
Through the plaquette singlet resonance, any snapshot of PSS (DPS)
can turn into another without any explicit transition.
PSS and DPS share DP and RD as their snapshots.
Thus, the extended variational wave function (\ref{E:PSS}) is
expected to erase the artificial first-order transition lines
(thin solid lines), reducing the only discontinuity wall AB to a
point.
The point A belongs to the same universality class as the
spin-$\frac{1}{2}$ Heisenberg chain \cite{Y16128}.
We are all the more convinced of the immediate gap formation with $r$
moving away from $0$.
On the other hand, neither PSS nor DPS includes both OPLD and SH and
therefore the two critical lines in the out-of-phase-leg region
survive against the plaquette singlet formation.

\begin{figure}
\centerline
{\mbox{\quad\qquad\qquad\psfig{figure=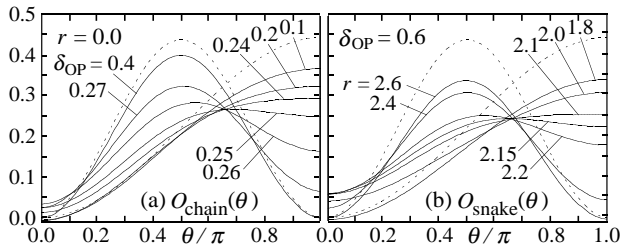,width=100mm,angle=0}}}
\vspace*{-15mm}
\caption{Quantum Monte Carlo estimates of the generalized string
         order parameter defined on the linear-chain (a) and snake
         (b) paths.
         The dashed lines represent the analytic calculations for the
         AKLT VBS ($\frac{4}{9}\sin^2\frac{\theta}{2}$) and 
         decoupled dimers ($\frac{4}{9}\sin^2\theta$).
         $\delta_{\rm OP}$ runs on the line of $r=0.0$ in (a), while
         $r$ runs on the line of $\delta_{\rm OP}=0.6$ in (b).}
\label{F:string}
\end{figure}

   The generalized string order parameter \cite{O7469}
$
   O(\theta)
    =\lim_{|i-j|\rightarrow\infty}
     \langle
     S_i^z
     \prod_{l=i}^{j-1}\exp[{\rm i}\theta S_l^z]
     S_j^z
     \rangle
$
is also useful in characterizing the ground state.
$O(\theta)$ distinguishes between one-dimensional VBS states by its
$\theta$ dependence \cite{Y3603}.
Hence, measuring it on the linear-chain and snake paths, we can
detect the transitions between DH, OPLD, SH, and RD, as is shown in
Fig. \ref{F:string}.
If we specify the transition through a change of the $\theta$
dependence in the vicinity of $\theta=\pi$, that is, the change from
the convex curve to the concave one, we obtain the transition points
$(\delta_{\rm OP},r)=(0.245,0)$ and $(0.6,2.125)$, which are in good
agreement with the numerical findings in Fig. \ref{F:PhD}(b).
Whatever path we take for $O(\theta)$, its peak never sits on
$\theta=\pi$ in the in-phase-leg region, suggesting that we can not
observe the Haldane state of any kind there.
The plaquette singlet formation can instead be visualized by
extending $O(\theta)$ to ladders as
$
 \lim_{|i-j|\rightarrow\infty}
 \langle
  S_{1,i}^z S_{2,i}^z
  \prod_{l=i}^{j-1}\exp[{\rm i}\theta(S_{1,l}^z+S_{2,l}^z)]
  S_{1,j}^z S_{2,j}^z
 \rangle
$
\cite{T}.
\vspace*{-10mm}
\vspace*{-1mm}

   We summarize our rich harvest as the plaquette singlet formation
and the re-entrant quantum phase transition via the snake Haldane
state in staggered spin-$1$ ladders.
The re-entrant phase boundary is peculiar to spin-$1$ or possibly
integer-spin ladders.
Half-odd-integer-spin chains are generically critical at the
translationally symmetric point $\delta=0$.
Therefore, the critical line can not exhibit the initial re-entrant
behavior in the $r$-$\delta$ phase diagram, because it should be
symmetric for $\pm\delta$.
Chemical tuning of bond-alternating critical chain compounds
\cite{H1312}, as well as the organic-radical-based materials research
\cite{K1008}, is strongly encouraged.

We are grateful to H. Takayama and K. Okamoto for fruitful
discussions.
This work was supported by the Japanese Ministry of Education,
Science, Sports, and Culture.
The numerical calculation was done using the facility of the
Supercomputer Center, Institute for Solid State Physics, University
of Tokyo.

\widetext

\begin{references}

\bibitem{H464}
   F. D. M. Haldane:
      Phys. Lett. {\bf 93A} (1983) 464;
      Phys. Rev. Lett. {\bf 50} (1983) 1153.

\bibitem{R945}
   J. P. Renard, M. Verdaguer, L. P. Regnault, W. A. C. Erkelens,
   J. Rossat-Mignod and W. G. Stirling:
      Europhys. Lett. {\bf 3} (1987) 945.

\bibitem{A799}
   I. Affleck, T. Kennedy, E. H. Lieb and H. Tasaki:
      Phys. Rev. Lett. {\bf 59} (1987) 799;
      Commun. Math. Phys. {\bf 115} (1988) 477.

\bibitem{H3651}
   M. Hase, I. Terasaki and K. Uchinokura:
      Phys. Rev. Lett. {\bf 70} (1993) 3651.

\bibitem{O1984}
   M. Oshikawa, M. Yamanaka and I. Affleck:
      Phys. Rev. Lett. {\bf 78} (1997) 1984.

\bibitem{Y14008}
   S. Yamamoto and T. Fukui:
      Phys. Rev. B {\bf 57} (1998) R14008.

\bibitem{D5744}
   E. Dagotto, J. Riera and D. Scalapino:
      Phys. Rev. B {\bf 45} (1992) R5744.

\bibitem{A3463}
   M. Azuma, Z. Hiroi, M. Takano, K. Ishida and Y. Kitaoka:
      Phys. Rev. Lett. {\bf 73} (1994) 3463.

\bibitem{U2764}
   M. Uehara, T. Nagata, J. Akimitsu, H. Takahashi,
   N. M$\hat{\mbox o}$ri and K. Kinoshita:
      J. Phys. Soc. Jpn. {\bf 65} (1996) 2764.

\bibitem{D618}
   E. Dagotto and T. M. Rice:
      Science {\bf 271} (1996) 618.

\bibitem{H12649}
   C. A. Hayward, D. Poilblanc and L. P. L\'evy:
      Phys. Rev. B {\bf 54} (1996) R12649.

\bibitem{L100402}
   C. P. Landee, M. M. Turnbull, C. Galeriu, J. Giantsidis and
   F. M. Woodward:
      Phys. Rev. B {\bf 63} (2001) R100402.

\bibitem{K1008}
   K. Katoh, Y. Hosokoshi, K. Inoue and T. Goto:
      J. Phys. Soc. Jpn. {\bf 69} (2000) 1008.

\bibitem{H12924}
   Y. Hosokoshi, Y. Nakazawa, K. Inoue, K. Takizawa, H. Nakano,
   M. Takahashi and T. Goto:
      Phys. Rev. B {\bf 60} (1999) 12924.

\bibitem{H}
   Y. Hosokoshi, K. Katoh, Y. Nakazawa, H. Nakano and K. Inoue:
      to be published in J. Am. Chem. Soc..

\bibitem{S3299}
   G. Sierra:
      J. Phys. A: Math. Gen. {\bf 29} (1996) 3299.

\bibitem{A397}
   I. Affleck:
      Nucl. Phys. {\bf B257} (1985) 397;
                  {\bf B265} (1986) 409.

\bibitem{E2626}
   R. S. Eccleston, T. Barnes, J. Brody and J. W. Johnson:
      Phys. Rev. Lett. {\bf 73} (1994) 2626.

\bibitem{M3443}
   M. A. Martin-Delgado, R. Shankar and G. Sierra:
      Phys. Rev. Lett. {\bf 77} (1996) 3443.

\bibitem{F398}
   T. Fukui and N. Kawakami:
      Phys. Rev. B {\bf 57} (1998) 398.

\bibitem{L343}
   A. Langari, M. Abolfath and M. A. Martin-Delgado:
      Phys. Rev. B {\bf 61} (2000) 343.

\bibitem{P4393}
   A. Parola and S. Sorella and Q. F. Zhong:
      Phys. Rev. Lett. {\bf 71} (1993) 4393.

\bibitem{W886}
   S. R. White, R. M. Noack and D. J. Scalapino:
      Phys. Rev. Lett. {\bf 73} (1994) 886.

\bibitem{F3714}
   B. Frischmuth, B. Ammon and M. Troyer:
      Phys. Rev. B {\bf 54} (1996) R3714.

\bibitem{T}
   S. Todo, M. Matsumoto, C. Yasuda and H. Takayama:
      cond-mat/0107115.

\bibitem{A1196}
   P. W. Anderson:
      Science {\bf 235} (1987) 1196.

\bibitem{R445}
   T. M. Rice, S. Gopalan and M. Sigrist:
      Europhys. Lett. {\bf 23} (1993) 445.

\bibitem{S4053}
   T. Sakai and S. Yamamoto:
      Phys. Rev. B {\bf 60} (1999) 4053.

\bibitem{N561}
   M. P. Nightingale:
      Physica A {\bf 83} (1976) 561.

\bibitem{N5773}
   K. Nomura and K. Okamoto:
      J. Phys. A: Math. Gen. {\bf 27} (1994) 5773.

\bibitem{Y16128}
   S. Yamamoto:
      Phys. Rev. B {\bf 51} (1995) 16128;
                   {\bf 52} (1995) 10170.

\bibitem{K285}
   A. Kitazawa:
      J. Phys. A: Math. Gen. {\bf 30} (1997) L285.

\bibitem{S2484}
   R. R. P. Singh, M. P. Gelfand and D. A. Huse:
      Phys. Rev. Lett. {\bf 61} (1988) 2484.

\bibitem{K6133}
   A. Koga and N. Kawakami:
      Phys. Rev. {\bf 61} (2000) 6133.

\bibitem{G13}
   A. J. Guttmann:
      {\it Phase Transitions and Critical Phenomena},
      ed. C. Domb and J. L. Lebowitz
      (Academic, New York, 1989) Vol. 13.

\bibitem{Y3603}
   S. Yamamoto:
      Phys. Rev. B {\bf 55} (1997) 3603.

\bibitem{O7469}
   M. Oshikawa:
      J. Phys.: Condens. Matter {\bf 4} (1992) 7469.

\bibitem{H1312}
   M. Hagiwara, Y. Narumi, K. Kindo, H. Nakano, R. Sato and
   M. Takahashi:
      Phys. Rev. Lett. {\bf 80} (1998) 1312.

\end{references}
\end{document}